\documentstyle[12pt,a4]{article}
\begin{document}
\begin{center}
{\Large \bf COMPARATIVE ANALYSIS OF MULTIPLICITY DISTRIBUTIONS IN INELASTIC
PROCESSES FOR DIFFERENT COLLIDING PARTICLES AND NUCLEI}

{\large I.M.~Dremin$^{1}$, 
V.A.~Nechitailo$^{1}$, M.~Biyajima$^{2}$, N.~Suzuki$^{3}$}

\vspace{2mm}
{\it ${}^{1}$P.N.~Lebedev Physical Institute, Moscow, Russia}

{\it $^{2}$Department of Physics, Faculty of Liberal Arts,\\
Shinshu University, Matsumoto 390, Japan }

{\it $^{3}$Matsusho Gakuen Junior College, Matsumoto 390-12, Japan }

\end{center}

\begin{abstract}
Theoretical prediction of oscillations of cumulant moments of parton
multiplicity distributions inside a jet supported by experimental data in some
multiple production processes asks for analysis of the phenomenon for the
whole set of available reactions. We have found out that the oscillations
persist in any kind of processes and increase for particles with more
complicated structure i.e. in the order of $ee, eh, hh, hA, AA$. The effect
is not strongly dependent on the available phase space.

Theoretical values of moments for quark and gluon jets up to 5th rank are
shown. Zeros of the truncated generating function and singularities of the
total generating function are discussed.
\end{abstract}
\section{Introduction}
Multiplicity distributions and their moments contain in the integrated form
all the information about a multiple production process. At the same time,
their analysis is rather simple since it does not ask for multidimensional plots
 and its results are easily interpreted. It is especially important that QCD
provides quite detailed predictions on the behavior of parton multiplicity
distributions. Even though the direct comparison with experimental data is
impossible without Monte Carlo simulations with definite assumptions about
hadronization, qualitative features of the predicted (parton) and measured
(hadron) distributions are so similar that one is tempted to say that
experiment supports the predictions. Further support is given to the local
parton-hadron duality hypothesis as well.

The most impressive prediction of QCD is the minimum \cite{1} of cumulant
moments (or their ratio to factorial moments) at the rank $q \approx 5$ with
a negative value at the minimum and subsequent oscillations \cite{2} (for
review see \cite{3}). It concerns parton distributions. With the local
parton-hadron duality at work, one could hope for the similar behavior of
hadron distributions in gluon and quark jets i.e. primarily in $e^{+}e^{-}$
annihilation. It is astonishing that such features have been observed not
only in $e^{+}e^{-}$ but in $pp$ \cite{4}, $pA$ and $AA$ \cite{5} as well.
Here we complete the list of available reactions by analyzing H1 data on
$ep$ deep inelastic processes \cite{6} and show regularities which appear for
the whole set of the processes.

The similarity of patterns of cumulant moments oscillations in all the
processes reveals some general origin of dynamics, probably indicating their
underlying cascade nature. This similarity is especially surprising because
the shapes of multiplicity distributions for these processes differ drastically.
The minimum of cumulant moments at $q \approx 5$ becomes deeper for colliding
particles with more complicated structure i.e. in the order of them like
$e^{+}e^{-}, eh, hh, hA, AA$. This regularity has no physical interpretation
yet. It does not contradict, however, to results obtained by Monte Carlo
simulation schemes.

In view of the existing \cite{DELPHI} and forthcoming \cite{OPAL} attempts to
study multiplicity distributions inside gluon and quark jets separately
we provide our predictions for their cumulant moments up to the rank 5,
comparing them to DELPHI data \cite{DELPHI}.

We have also calculated the location of zeros of the truncated generating
function for $ep$ collisions and compared it with corresponding pattern in
other reactions. The possible scaling of zeros location and the nature of
the singularity of the total generating function are discussed. This is
appealing in view of the fact that asymptotically the singularity is located
in the vicinity of the point where moments are calculated as derivatives
of the generating function. The similarity to the partition function properties
in statistical machanics where they are related to phase transition problems
gives further impact to that study.

\section{Moments of multiplicity distributions}
We have analyzed new data of H1 Collaboration \cite{6} on multiplicity
distributions in deep inelastic scattering for different energy (W) intervals
in terms of moments. This method proved to be useful for $ee, pp$ \cite{4}
and $hA, AA$ \cite{5} processes. The cumulant moments are especially sensitive
to tiny details of distributions and reveal the intriguing oscillations.

Let us define the normalized moments of the multiplicity distribution $P_{n}(y)$
where $P_{n}$ denotes the probability of $n$-particle production process at
c.m.s. energy $s^{1/2}$ and $y=\ln (s^{1/2}/m)$ with $m \sim $1 GeV. After
measuring $P_{n}(y)$ one can calculate the factorial moments $F_{q}$ of the
rank $q$ as
\begin{equation}
F_{q}=\frac {\sum _{n=0}^{\infty}n(n-1)...(n-q+1)P_{n}(y)}{(\sum _{n=0}^{\infty}
nP_{n}(y))^{q}}       \label{1}
\end{equation}
and subsequently the cumulant moments $K_{q}$ by iterative procedure from the
set of relations
\begin{equation}
F_{q}=\sum _{n=0}^{q-1}C_{q-1}^{m}K_{q-m}F_{m}    \label{2}
\end{equation}
where
\begin{equation}
C_{q-1}^{m}=\frac {(q-1)!}{m!(q-m-1)!}     \label{3}
\end{equation}
are binomial coefficients. For all high energy experimental distributions
factorial moments are larger than 1 but cumulant moments oscillate. Usually,
both factorial and cumulant (in modulo) moments increase fast with their
rank $q$. Therefore, it is convenient to consider their ratio
\begin{equation}
H_{q}=\frac {K_{q}}{F_{q}}.      \label{4}
\end{equation}
It is especially so because QCD provides results just in terms of $H_{q}$
\cite{3} predicting specific oscillations as a function of $q$ with first
minimum located at
\begin{equation}
q_{min}=\frac {1}{h_{1}\gamma _{0}} + \frac{1}{2} \approx 5      \label{5}
\end{equation}
where $h_{1}=11/24$ is given by the integral over the kernel of QCD evolution
equations, and $\gamma _{0}^{2} = 6\alpha _{s}/\pi, \alpha _{s}$ is the QCD
"fine structure constant" ($\gamma _{0} \approx 0.48$ at $Z^{0}$-peak).
This estimate is obtained for parton distributions in gluon jets.
For quark jets it could shift to $q\approx 7$.

In Fig.1 we have plotted the dependence of ratios $H_{q}$ on their rank $q$
for various initial colliding partners ($ee, ep, pp, pA, AA$) at high energies.
All of them reveal the oscillations with first minimum located at $q \approx 5$
or nearby as predicted by QCD (for parton distributions in gluon jets!). The
mentioned regularity in the depth of first minimum is clearly seen. It becomes
deeper for those reactions where the internal structure of the colliding 
objects is more complicated. This regularity has been noticed in \cite{4,5} 
and new data on $ep$ collisions \cite{6} fits it very well.

The oscillations could be attributed to some common cascade nature of the
processes even though the present theoretical approach to $ee$ differs
drastically from $AA$ description ($s$-channel cascades versus $t$-channel ones).
Their periodicity and amplitude are even more intriguing. Anyway, no
conventional phenomenological (probability theory) distribution fits them.

The attempts to ascribe all the oscillations to the cut-off of
multiplicities at finite energy fail because it gives rise to  comparatively
small values of amplitudes
if maximal multiplicity is much larger than  the average one\cite{8}
and, even more important, predicts smaller amplitudes in $AA$ as compared to
other processes due to higher multiplicities in $AA$. It contradicts Fig.1.
Nevertheless various cut-offs are rather important provided they limit
substantially the available  multiplicity. The H1 data \cite{6} at the same
energy W but for different rapidity windows confirm this statement if plotted
in terms of $H_q$ in Fig.2. The smaller window one has chosen, the larger is
the amplitude of the oscillations with the first minimum shifted to higher
ranks.

As seen from eq. (5), the minimum location in QCD depends on coupling constant
and on the kernel of evolution equations where the vector nature of gluons is
of primary importance.

Its dependence on the QCD running coupling is such  that $q_{min}$ should
be slightly larger at higher energies because $\gamma _{0}$ (or $\alpha _{s}$)
decreases. This shift is rather hard to notice since the decrease of
$\gamma _{0}$ is not strong enough and, besides, the values of moments are only
available at integer ranks. Some qualitative indications can, however, be
obtained if one deals with data of the same collaboration. Otherwise, different
selection criteria are difficult to disentangle. This unique possibility is
provided by recent H1 data \cite{6}. Our analysis shows that QCD trends
are  rather weak and multiplicity cut-off effect conceals these trends.
In Fig.3 we plot the moments for different "excitation energies" (W) of the
proton calculated at the same maximal multiplicity equal to 18.
We see that the minimum tends to move to the left at higher W.
Therefore, cut-off effect is still strong enough  if maximal multiplicities
are not much larger than the average multiplicity.

Another chance to check QCD results appears when gluon and quark jets are
accurately isolated and considered separately as is done in refs.
\cite{OPAL,DELPHI},
for example. Then the widths and higher moments (apart average multiplicities
considered in \cite{OPAL}) of multiplicity distributions within jets can be
treated.
In Table I, we show the values of corresponding moments as given by fixed
coupling QCD equations. However the value of the coupling constant has been
chosen higher than at $Z^0$-peak to take into account its running during
the cascade evolution. According to the analytic results \cite{1}, the cumulant
moments depend rather strongly on the coupling constant. It has been confirmed
by varying its value in numerical estimates. The moments decrease for both
gluon and quark jets if coupling constant increases. However we have found that 
the ratios $H_q^G/H_q^Q$ are rather stable for different $\gamma_0$ being
approximately equal to 2 for $q=2$ and ranging from 4 to 5 for $q=3$.
One must remember however that there are two predicted numbers to be 
compared with experiment for each rank, not a single one.

As seen from Table I, the width of the multiplicity distribution for 
a quark jet is larger than that for a gluon jet. At present, there is DELPHI 
data where the values of the Negative Binomial Distribution parameter $k$ for 
multiplicity distributions in quark and gluon jets are shown in Tables 5 and 7 
of \cite{DELPHI}. Misleadingly, they are called dispersions in the titles to the
Tables even though the higher values of $k$ correspond to lower dispersions. 
They have been derived from the fit to the Negative Binomial 
Distribution using the DURHAM and JADE algorithms. The values of $k$ obtained 
according to JADE (Table 7 in \cite{DELPHI}) and DURHAM (Table 5 
in \cite{DELPHI}) algorithms differ drastically, probably due to the different 
angular coverage for those algorithms, and contradict to each 
other (especially, at low jet energies). Unfortunately, the statistical errors 
are shown only in the Tables. They are large but one can suspect that the
systematical errors (not shown) can be even larger due to $b$ and $c$ quark
jets not separated. To compare theory to the data such a separation should be 
done. Otherwise from the data considered boldly, one would conclude that 
there is a contradiction between theory and experiment. Better understanding
of these systematic effects is needed in order to make a fair comparison
between data and theory in this subject.
Higher moments ask for better jet separation and statistics also.

\section{The generating function}

Factorial and cumulant moments are calculated from experimental multiplicity
distributions according to eqs. (1)-(3). For theorists, it is more convenient
to deal with the generating function
\begin{equation}
G(z,y)=\sum _{n=0}^{\infty}(1+z)^{n}P_{n}(y)=\sum _{q=0}^{\infty}\frac {F_{q}}
{q!}z^{q}\langle n\rangle ^{q},      \label{6}
\end{equation}
where $\langle n\rangle =\sum _{n=0}^{\infty}nP_{n}$ is the average
multiplicity. If $G(z,y)$ is known, the factorial and cumulant moments are
given by its derivatives
\begin{equation}
F_{q}=\frac {1}{\langle n\rangle ^{q}}\left. \frac {d^{q}G(z,y)}{dz^{q}}
\right\vert _{z=0},     \label{7}
\end{equation}
\begin{equation}
K_{q}=\frac {1}{\langle n\rangle ^{q}}\left. \frac {d^{q}\ln G(z,y)}{dz^{q}}
\right\vert _{z=0}.      \label{8}
\end{equation}
Therefore it is important to know how close the singularity of $G(z,y)$ in
$z$-plane is located to the point $z$=0 where the moments are calculated. The
nature of the singularity is also important to compare it to the generating
functions of phenomenological distributions and to look for some statistical
analogies with their speculations about phase transitions (see \cite{3}).
The analytic structure of the generating function in gluodynamics is
determined by the integro-differential equation
\begin{equation}
\frac {dG(z,y)}{dy}=\int _{0}^{1}dx[\frac {1}{x}-(1-x)(2-x(1-x))]\gamma _{0}^{2}
[G(z,y+\ln x)G(z,y+\ln (1-x))-G(z,y)].     \label{9}
\end{equation}
In QCD it is replaced by a set of two equations. Until now, nobody was able
to get the exact solution of this equation in terms of $G(z,y)$. It has been
solved in terms of the moments exactly for fixed coupling constant and in
higher order approximations for running coupling constant (see \cite{3}). It
is well known \cite{9} that in the double-logarithmic (DLA) approximation the
equation reduces to
\begin{equation}
\frac {d^{2}\ln G(z,y)}{dy^{2}}=\gamma _{0}^{2}[G(z,y)-1].     \label{10}
\end{equation}
It is quite clear that the leading singularity governed by eq. (10) can be
written as
\begin{equation}
G(z,y) \propto (z-z_{0}(y))^{-2}      \label{11}
\end{equation}
because differentiating $\ln G$ twice produces the second order pole in
the left hand side that asks for the same singularity in
the right hand side. Since KNO-scaling is  valid in this approximation,
the factorial moments do not depend asymptotically on $y$, and the
generating function (6) depends on the product of $z\langle n\rangle $
i.e. $z_{0}(y)=\kappa /\langle n\rangle $. The constant $\kappa $ is
equal to 2.552 \cite{9}.

The eq. (11) reminds the negative binomial distribution with its parameter
$k_{NBD}=2$ that corresponds to the extremely wide multiplicity distribution. In
the last case, the constant $\kappa $ would, however, be equal to 2 instead
of 2.552. More complete formula for $G(z,y)$ near the singularity looks in
DLA like
\begin{equation}
G(z,y)=\frac {2z_{0}^{2}}{(z-z_{0})^{2}}+\frac {2z_{0}}{z-z_{0}}-\frac {2}{3}
\ln \frac {z_{0}-z}{z_{0}}+O(1).       \label{12}
\end{equation}
This formula has rather heuristic than practical interest because experimental
distributions are far from being fitted by the negative binomial distribution
with $k_{NBD}=2$. It shows, however, that $z=0$ is still quite far from the
singularity (which moves to $z=0$ asymptotically!) in the sense that the pole
contributions cancel at this point in $G(z,y)$ itself. There is no cancellation
in its derivatives which increase like factorials due to the leading
singularity.

The attempts to go beyond DLA are connected with Taylor series expansion of $G$
(see \cite{3}) so that the equation (10) gets in the right hand side additional
terms proportional to derivatives of $G(z,y)$. Each derivative has stronger
singularity than the function itself and, therefore, such an approach can not be
used near the singular points. It demonstrates also why DLA does not work in
practice. One has to come back to the original equation (9) which has not been
solved yet.

From the experimental side, there is another way to approach the problem. One
can study the behavior of zeros of the truncated generating function
$G_{tr}(z,y)$ which is defined by
\begin{equation}
G_{tr}(z,y)=\sum _{n=0}^{n_{0}}(1+z)^{n}P_{n}(y)     \label{13}
\end{equation}
and differs from the total generating function by the sum being truncated at
some value of $n=n_{0}$ that happens inevitably in experiment due to finite
energy. In principle, one can truncate at different values $n_{0}$ for a
given energy $y$ up to the maximum available multiplicity $n_{max}$. The
truncated generating function is a polynomial of order $n_{0}$ with $n_{0}$
complex conjugate zeros in the complex $z$-plane. It has no singularity at
finite $z$. The positions of zeros have been studied in several papers
(e.g. \cite{5,10,11}). It has been shown that at large $n_{0}$ their pattern
reminds the circle. We have analyzed DESY data and have found that the similar
regularity appears there as well with some additional substructure in the
left hand side of the circle due to low values of cut-off $n_{0}$. It can be
seen in Fig.4. One can compare our findings with previously published results
\cite{5,11,12}.

With $n_{0}$ increasing, zeros tend to the real axis and their limit at
$n_{0}\rightarrow \infty $ should coincide with the singularity of the
generating function. The relation of the density of zeros and of their
approach to the real axis to the singularity nature has been discussed in
\cite{12}. This kind of analysis of experimental results has not been done
yet. It requires special study and we hope to do it later and publish
elsewhere.

\section{Conclusions}

In previous papers \cite{4,5,10,11} the analysis of the moments of multiplicity
distributions and of zeros of the truncated generating functions has been done
in $ee, pp, pA, AA$ collisions. We complete this list of reactions doing the
same for deep inelastic $ep$ collisions using recently published H1 data. We
find out that this reaction fits quite well the general regularities observed
previously for oscillations of the moments and zeros of the truncated generating
function. In particular, the depth of the first minimum of oscillations is
intermediate between $ee$ and $pp$ minima supporting earlier conclusion about
its dependence on the structure of the colliding objects.

Some predictions for quark and gluon jets are discussed in connection with
recent experimental data and forthcoming attempts to measure jet moments.

Also, we discuss and try to attract attention to the problem of singularities
of the generating function which is still waiting for its solution. Our
findings are encouraging in that respect.

\section{Acknowledgment}
I.D. is grateful to J. Drees, J. Fuster, K. Hamacher, S. Marti i Garcia
for discussions concerning DELPHI data.
This work was supported in part by Russian Fund for Basic Research (grant
96-02-16347), by INTAS (grant 93-79) and by JSPS.

\newpage

\noindent {\Large\bf Figure Captions}
\begin{description}
\item[Fig.1] The ratio $H_q$ for $e^+e^-$, $e$p, p$\bar{\rm p}$, 
              hA and AA collisions.    (See text for more details).
\item[Fig.2]  The ratio $H_q$ for $e$p collisions at $150<W< 185$ GeV
              for different pseudorapidity intervals.
\item[Fig.3] The ratio $H_q$ for $e$p collisions for different W
             and fixed maximum multiplicity equal to 18.
\item[Fig.4] The zeros of the truncated generating function $G_{tr}(z)$ 
             for $e$p collisions at $150<W< 185$ GeV.
\end{description}

\newpage
\vspace*{1cm}
\begin{table}
\hspace*{4cm}\caption{$r = 1.8; n_F = 4; \gamma_0 = 0.648; \gamma = 0.476.$}
\vspace{2mm} \label{Tab2}
\begin{tabular}{|c|c|c|c|c|c|c|}
\hline
q &  $F^G_q$  & $F^Q_q$  & $K^G_q$ & $K^Q_q$ & $H^G_q$ & $H^Q_q$ \\
\hline
2 & 1.14 &  1.32  &  0.14 &  0.32 &   0.12 & 0.24 \\
\hline
3 & 1.45 & 2.14 &   2.9 10$^{-2}$ &  0.18 & 2.0 10$^{-2}$ & 8.5 10$^{-2}$\\
\hline
4 & 2.02 & 4.06 & 5.1 10$^{-3}$ &  0.12 &   2.5 10$^{-3}$ & 2.9 10$^{-2}$ \\
\hline
5 &3.05&8.76&-4.2 10$^{-4}$ &6.5 10$^{-2}$ &-1.4 10$^{-4}$ & 7.4 10$^{-3}$\\
\hline \end{tabular} \end{table}


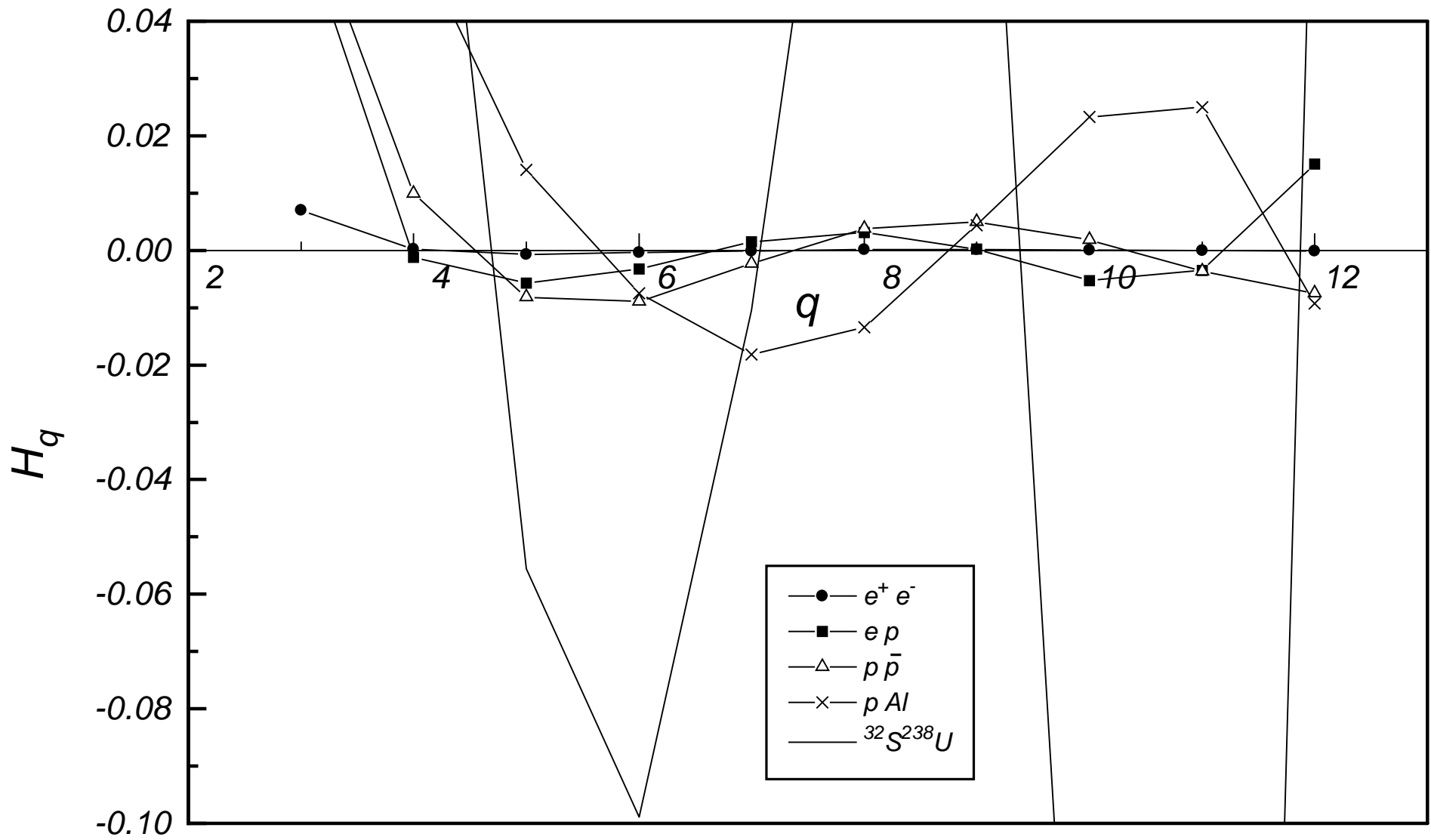
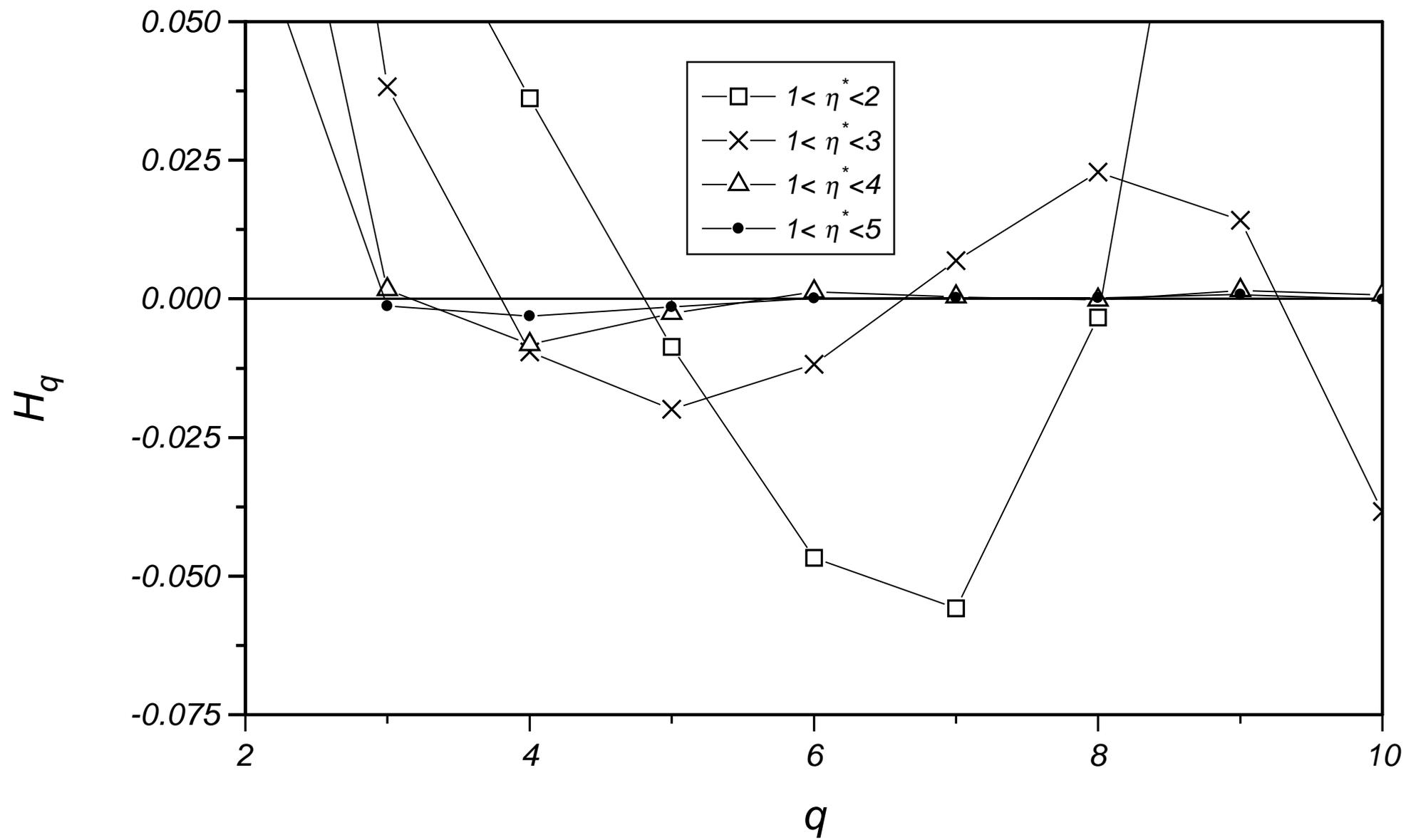
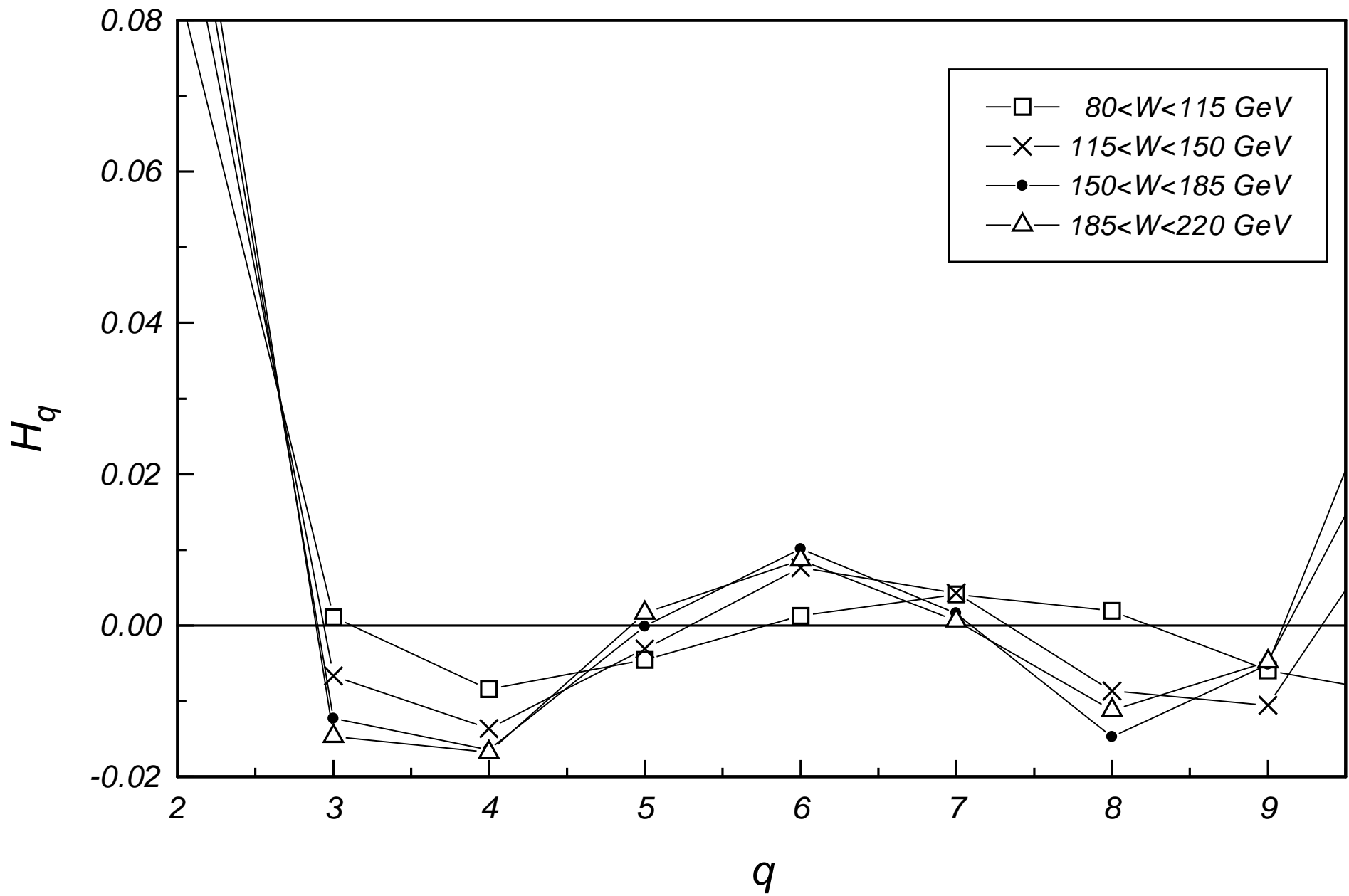
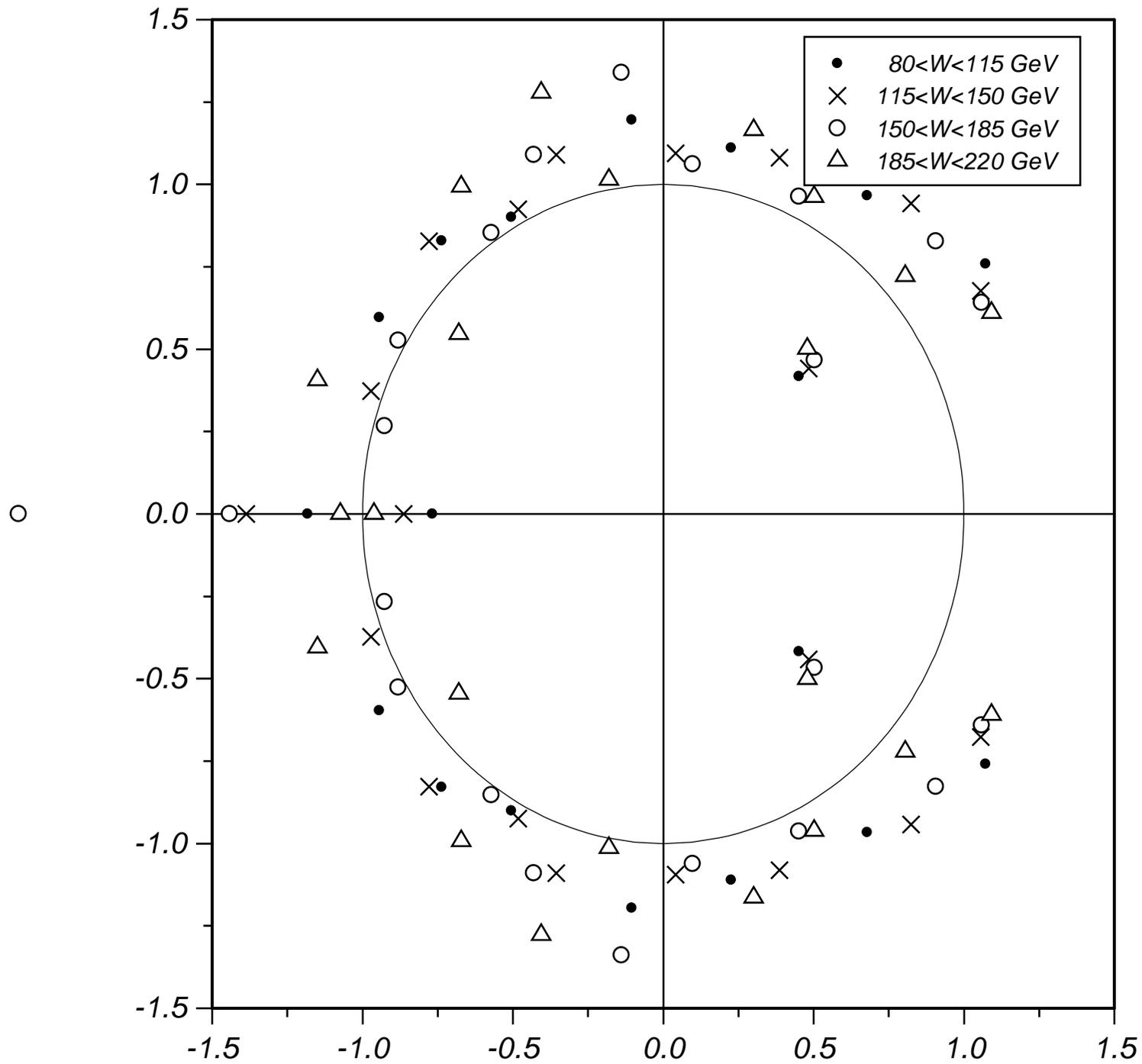
\end{document}